# Geometrical Interpretation of the Master Theorem for Divide-and-conquer Recurrences

Simant Dube[*]

Department of Computer Science and Engineering
Indian Institute of Technology
Kanpur 208016, India
Email: simantd (at) iitk.ac.in

**Abstract.** We provide geometrical interpretation of the Master Theorem to solve divide-and-conquer recurrences. We show how different cases of the recurrences correspond to different kinds of fractal images. Fractal dimension and Hausdorff measure are shown to be closely related to the solution of such recurrences.

**Keywords:** Divide-and-conquer recurrences, the Master Theorem, Fractal Geometry, Hausdorff Dimension, Hausdorff Measure, Fractal Dimension.

## 1 Introduction

Divide-and-conquer recurrences often arise in computer science as many important problems can be solved using divide-and-conquer approach in which a problem is divided into smaller subproblems which are solved independently and then their results are processed and combined to give final results. A classic example of this divide-and-conquer approach is Fast Fourier Transform (FFT) in which a problem of size $n$ is divided into two subproblems each of size $n/2$ and results of which are combined in linear time. The divide-and-conquer recurrence for FFT is

$$T(n) = 2T(n/2) + O(n)$$

which has the solution $T(n) = O(n \log n)$.

[*] Supported by a research fellowship from Govt. of India

It has been well known how to solve such divide-and-conquer recurrences, see [3] for the Master Theorem, and its various generalizations as in [1, 5, 8].

In this paper, we present a new geometrical way of looking at the Master Theorem. We make use of Fractal Geometry which is mathematical study of self-similar objects [2, 6]. It is natural to see the connection between fractals, which can be often created using a recursive subdivision process, and recursive algorithms such as FFT. For past related work, see [4]. However in this past work, the treatment of the subject and proofs of results are complex as one worked with infinite resolution images. It also lacks geometrical intuition and emphasis on the most common type of divide-and-conquer recurrences which are handled by the Master Theorem.

This paper derives the results for such common recurrences in a simpler and expository manner. The primary difference in approaches of this paper and of [4] is that we work with finite recursion trees and corresponding finite resolution images here to get the results in a simpler manner.

This link between fractals and the Master Theorem is remarkable as it gives simpler, intuitive understanding of the proof of the Master Theorem. Goal of this paper is to bridge this gap between divide-and-conquer algorithms and fractal geometry by providing this intuition to the reader. We geometrically illustrate each case of the Master Theorem and show how it is strongly related to the notions of Fractal dimension and Hausdorff measure.

Our main idea is to construct an Iterated Function System (IFS) [2] consisting of contractive transformations where each transformation corresponds to a recursive call to solve a subproblem. The recursions of the algorithm correspond to the iterations of the IFS. The time complexity of the algorithm is determined by the number of leaves in the recursion tree, which is identical to the number of $\epsilon$-balls needed to cover the fractal image in the computation of the Fractal dimension. The additional computation overhead of dividing the problem into subproblems and combining their results is also captured by a fractal image.

We conclude the paper by showing how a theorem on Fractal dimension proved in [7] is closely related to the Master Theorem.

Throughout the paper, we emphasize intuition by providing several examples which are illustrated in detail in figures.

## 2 Preliminaries

### 2.1 Master Theorem

We state the Master Theorem.

Let $a \geq 1$ and $b \geq 1$ be constants, let $f(n)$ be a function and let $T(n)$ be defined on the nonnegative integers by the recurrence

$$T(n) = aT(n/b) + f(n),$$

where we interpret $n/b$ to mean either $\lfloor n/b \rfloor$ or $\lceil n/b \rceil$. Then $T(n)$ can be bounded asymptotically as follows.

1. If $f(n) = O(n^{\log_b a - \epsilon})$ for some constant $\epsilon > 0$, then $T(n) = \Theta(n^{\log_b a})$.
2. If $f(n) = \Theta(n^{\log_b a})$, then $T(n) = \Theta(n^{\log_b a} \log n)$.
3. If $f(n) = \Omega(n^{\log_b a + \epsilon})$ for some constant $\epsilon > 0$, and if $af(n/b) \leq cf(n)$ for some constant $c < 1$ and all sufficiently large $n$, then $T(n) = \Theta(f(n))$.

For proof, see [3]. The three cases are needed as one has to consider if the overhead term $f(n)$ dominates the computation or the recursive term $aT(n/b)$ does. One term dominates the other when it is polynomially larger than the other. Note that for case 3 the theorem is overstated, as the regularity condition $af(n/b) \leq cf(n)$ for some constant $c < 1$ implies $f(n) = \Omega(n^{\log_b a + \epsilon})$ for some constant $\epsilon > 0$.

### 2.2 Hausdorff Dimension

For background on Hausdorff dimension see [2]. In 1, 2 or 3 dimensions, we have concept of length, area and volume, respectively, as a measure of size of an object. We can obtain this measure by covering the object by balls of diameter less than $\delta$ and summing up their sizes as defined by diameter raised to power $D$ where $D$, an integer, is the topological dimension. See Figure 1 (a).

Hausdorff measure generalizes this to non-integral dimension by considering *real* values of $D$ and examining how the measure would change as $D$ varies.

*Example 1.* The filled unit square $[0,1]^2$ has infinite 1-dimensional measure (i.e. informally, length) but zero 3-dimensional measure (i.e. informally, volume). In fact, when only $D = 2$ we will get finite measure when we sum up sizes of all the $\delta$-balls. For any real $D < 2$ we will have infinite measure and for any real $D > 2$ we will have zero measure, thereby indicating that $D = 2$ is the dimension of $[0,1]^2$. See Figure 1 (b) and (c).

This is true in general when we are trying to measure the size of a fractal object. If $D$ is less than the dimension of the object, then we should get infinite $D$-dimensional measure, whereas if $D$ is larger, then we should get zero value.

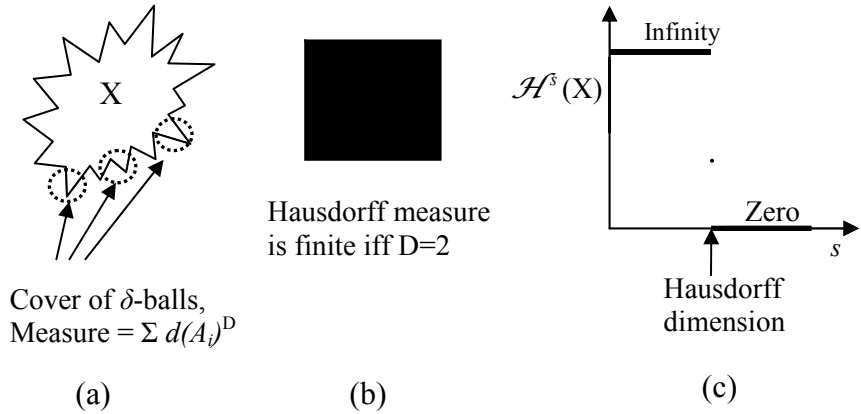

**Fig. 1.** Notions of Hausdorff Measure and Hausdorff Dimension.

Let $X$ be a bounded and closed (compact) subset of metric space $(\mathbb{R}^m, d)$ where $m$ is a positive integer and $d$ is the Euclidean metric. The *diameter* $d(X)$ is defined by

$$d(X) = \sup\{d(x,y) : x, y \in X\}.$$

Let $\delta > 0$. A (countable) family $\{A_i\}_{i \in \mathbb{N}}$ is a $\delta$-cover of $X$ if $(\forall i) d(A_i) < \delta$ and $X \subseteq \bigcup A_i$. For $s \geq 0$, define

$$\mathcal{H}^s_\delta(X) = \inf\{\sum_{i \in \mathbb{N}} d(A_i)^s : \{A_i\} \text{ is a } \delta\text{-cover of } X\}.$$

As $\delta$ decreases, there are fewer $\delta$-covers of $X$, and therefore $\mathcal{H}^s_\delta$ is non-decreasing and the value

$$\mathcal{H}^s(X) = \lim_{\delta \to 0} \mathcal{H}^s_\delta(X)$$

is well defined and may be infinite. $\mathcal{H}^s(X)$ is called the *s-dimensional Hausdorff measure* of $X$. It can be shown that if for some $s \geq 0$, $\mathcal{H}^s(X)$ is finite then for all $t > s$, $\mathcal{H}^t(X) = 0$. There will exist a unique real number $s \geq 0$ such that $\mathcal{H}^t(X) = \infty$ for $t < s$ and $\mathcal{H}^t(X) = 0$ for $t > s$, see Figure 1 (c). This point is called *Hausdorff dimension* of $X$:

$$\dim_H(X) = \inf\{s \geq 0 : \mathcal{H}^s(X) = 0\}.$$

## 2.3 Iterated Function Systems

A collection of $k$ contractive transformations $w_i : \mathbb{R}^m \to \mathbb{R}^m$ is called an *iterated function system* (IFS) [2]. A contractive transformation $w$ brings every pair of points closer by a contractivity factor $s < 1$. That is, $d(w(x), w(y)) < sd(x, y)$. Starting with an arbitrary compact subset $Y \subseteq \mathbb{R}^m$ and by applying iteratively these transformations and taking their union,

$$W(Y) = \bigcup_{i=1}^{k} w_i(Y)$$

one can obtain a self-similar fractal object $X$ as the attractor and fixed point of $W$

$$X = W(X) = \lim_{j \to \infty} W^{(j)}(Y).$$

In this article, IFS will be *non-overlapping* as defined in [2]. Intuitively, for any $i, j, i \neq j$, $w_i(X)$ and $w_j(X)$ do not overlap. They may

be just-touching at their boundary points though. *Fractal dimension* of $X$ is given by

$$\dim_F(X) = \lim_{\epsilon \to 0} \frac{\log \mathcal{N}(X, \epsilon)}{\log \frac{1}{\epsilon}}$$

where $\mathcal{N}(X, \epsilon)$ is smallest number of closed balls of radius $\epsilon > 0$ needed to cover $X$. For non-overlapping IFS,

$$\dim_F(X) = \dim_H(X)$$

and therefore we will use Hausdorff and Fractal dimensions interchangeably. This equivalence allows fast computation of Hausdorff dimension of $X$ as the solution of $D$ in

$$\sum_{i=1}^{k} s_i^D = 1 \qquad (1)$$

where $s_i$ is the contractivity factor of $w_i$, see [2].

## 3 Fractal Interpretation of the Recursive Term $aT(n/b)$

Consider the term $aT(n/b)$ in the recurrence equation. Assume $n$ is exact power of $b$. The problem of size $n$ is divided into $a$ subproblems each of size $n/b$. $O(1)$ computation occurs at the leaf nodes of the recursion tree $R$. $R$ is an $a$-ary tree of height $\log_b n$ and has $n^{\log_b a}$ leaves, see Figure 2 (a). Computation of $aT(n/b)$ is exactly determined by the total number of leaves.

We build its IFS analog. It is a non-overlapping IFS whose attractor $X$ is composed of union of its $a$ smaller copies obtained by applying $a$ many transformations each of contractivity factor $1/b$

$$X = \bigcup_{i=1}^{a} w_i(X) = W(X)$$

From Equation (1), the Hausdorff dimension of $X$ is given by solution of $D$ in

$$\sum_{i=1}^{a} (1/b)^D = 1$$

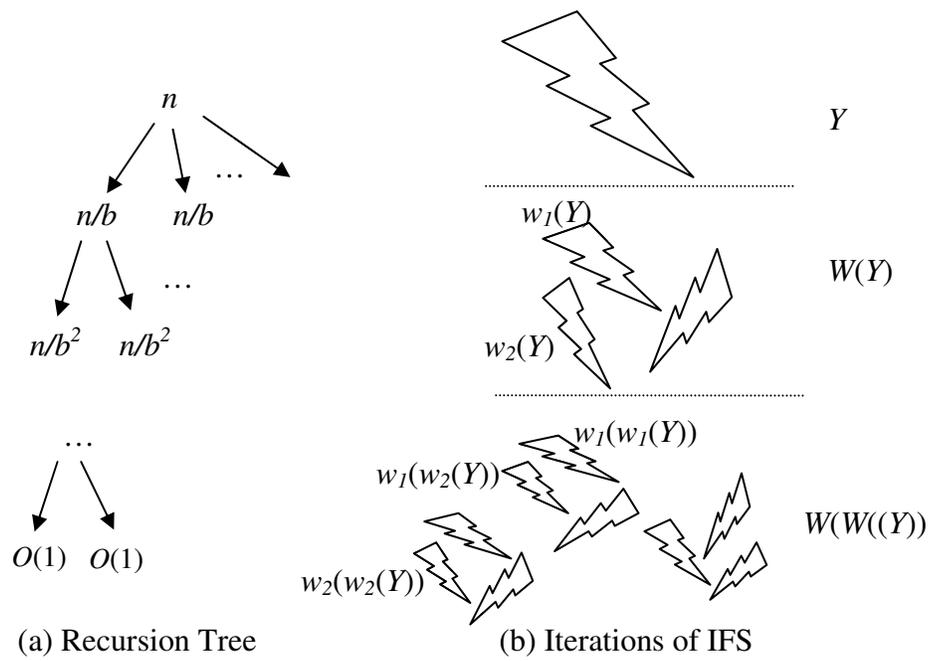

**Fig. 2.** Recursion tree for the recurrence term $aT(n/b)$. Iterations of IFS correspond to the recursions of the divide-and-conquer algorithm.

which gives $\dim_H(X) = \log_b a$. Starting with some set $Y$ of diameter 1, an iteration of $W$ creates $a$ smaller copies of the previous set as shown in Figure 2 (b). After $\log_b n$ iterations, we have the set $Y' = W^{(\log_b n)}(Y)$ which can be covered by $\mathcal{N}(Y', 1/n)$ many balls of radius $1/n$. Clearly, there will be as many such balls as the leaves on the recursion tree $R$ of the recurrence:

$$\mathcal{N}(Y', 1/n) = T(n) = \Theta(n^{\dim_H(X)}) = \Theta(n^{\dim_F(X)}). \tag{2}$$

Since we want the IFS to be non-overlapping, one has to show that such an IFS will exist. Consider the unit cube $[0,1]^m \subset \mathbb{R}^m$. It can be partitioned into $b^m$ many non-overlapping subcubes each of side length $1/b$. Just choose $m$ large enough so that $a \leq b^m$. Choose $a$ of these subcubes which will then determine $w_i$'s of the IFS mapping $[0,1]^m$ to these subcubes. To see why Equation (2) is true, compute $\mathcal{B}(X, 1/n)$ which is number of subcubes (boxes) intersecting $X$ in the box-counting algorithm to compute $\dim_F(X)$. The box-counting theorem, see [2], states that

$$\mathcal{N}(X, 1/n) = \Theta(\mathcal{B}(X, 1/n)).$$

*Example 2.* See Figure 3 for examples of fractal interpretation of recurrence term of the form $aT(n/2)$. The IFS which generate these fractal images use contractive quadrant affine transformations. These affine transformations are illustrated by showing how they map the unit square $[0,1]^2$ to quadrants. For sake of completeness, here are the transformations:

$$\begin{aligned}
w_1(x, y) &= (0.5x, 0.5y) \\
w_2(x, y) &= (0.5x + 0.5, 0.5y) \\
w_3(x, y) &= (0.5x, 0.5y + 0.5) \\
w_4(x, y) &= (0.5x + 0.5, 0.5y + 0.5) \\
w_5(x, y) &= (0.5x + 0.25, 0.5y + 0.5)
\end{aligned}$$

## 4 Fractal Interpretation of the Overhead Term $f(n)$

The second term $f(n)$ in the recurrence denotes the time taken to subdivide the problem and to combine the results of subproblems.

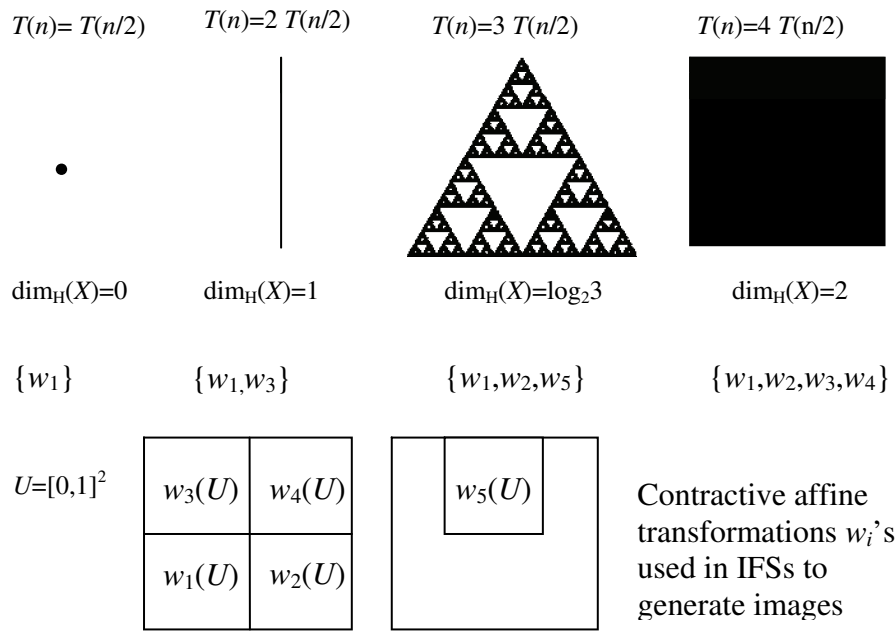

**Fig. 3.** Geometrical interpretation of the recurrence term of the form $aT(n/2)$, where $a \in \{1, 2, 3, 4\}$. The images have Hausdorff dimensions of $0, 1, \log_2 3$ and $4$, respectively. The IFS which generate these images are also shown.

Our trick is to consider $f(n)$ as another fractal image $C$ of dimension $p$ if $f(n) = \Theta(n^p)$. The IFS is generalized as follows:

$$W(Y) = (\bigcup_{i=1}^{a} w_i(Y)) \bigcup C$$

that is, we add $C$ at each iteration of the IFS. Such an IFS is also referred to as *Condensation IFS*, see [2].

After $j = \log_b n$ iterations, we have

$$Y' = W^{(j)}(Y) \cup C \cup W(C) \cup W^{(2)}(C) \cup \ldots \cup W^{(j-1)}(C).$$

If $Y$ and $C$ are of diameter 1, then it follows from the definition of $C$ that

$$\mathcal{N}(C, 1/n) = \Theta(n^p) = f(n),$$

and from the definition of $W$ that

$$\mathcal{N}(W^{(j-1)}(C), 1/n) = a^j f(n/b^j).$$

This and Equation (2) imply

$$\mathcal{N}(Y', 1/n) = \Theta(n^{\log_b a}) + \sum_{j=0}^{\log_b n - 1} a^j f(n/b^j) = T(n). \quad (3)$$

*Example 3.* Consider the recurrence where $f(n)$ is linear function:

$$T(n) = T(n/2) + \Theta(n).$$

Let $C$ be a fractal image with $\dim_H(C) = 1$. As we saw in the previous section, the first term $T(n/2)$ corresponds to an IFS with single transformation $w$ of contractivity factor $1/2$. For our illustration, we choose

$$w(x, y) = (0.5x, 0.5y)$$

and $C$ to be a tree like image as shown in Figure 4 and standing at point (1,0). Choose $Y = C$. We show first 3 iterations of the condensation IFS in Figure 4.

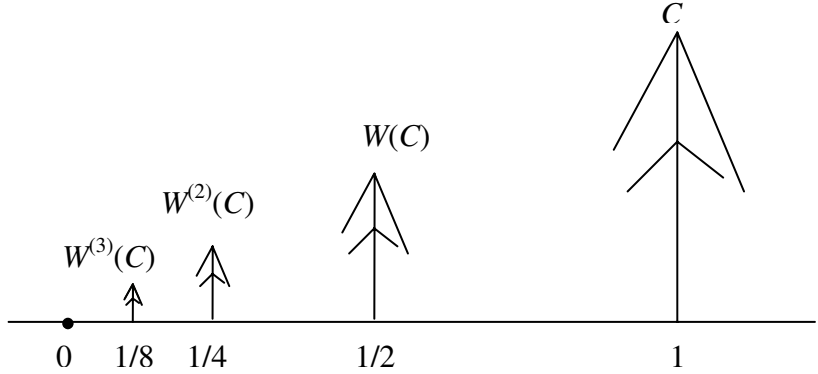

**Fig. 4.** First 3 iterations of the condensation IFS modeling the recurrence $T(n) = T(n/2) + \Theta(n)$. $C$ is a tree-like image with $\dim_H(C) = 1$ and models the $f(n) = \Theta(n)$ term. In this example, all computation is dominated by $f(n)$ term.

## 5 Fractal Interpretation of the Complete Recurrence

To complete the relationship of the recurrence with Fractal Geometry, we will need a result about Hausdorff dimension of attractors of Condensation IFS by Mauldin and Williams [7].

The theorem of Mauldin and Williams applies to more general IFS, referred to as graph directed IFS. In the case of condensation IFS, the graph consists of two nodes $u$ and $v$. Node $v$ represents the IFS for the term $aT(n/b)$ and its attractor $X_v$ has dimension $D_v = \log_b a$ as described in Section 3. The node $u$ represents the condensation image $C$ of dimension $D_u = p$, for the term $f(n)$, as described in Section 4. There are $a$ edges from $v$ to itself labeled by transformations $w_i, i \in \{1, 2, \ldots, a\}$. There is one edge from $u$ to $v$ labeled by identity transformation. See Figure 5.

Mauldin-Williams theorem has the following corollary.

**Theorem 1.** *Consider the condensation IFS in Figure 5.*

1. *The Hausdorff Dimension of the attractor $X$ of the condensation IFS is $D = \max(D_v, D_u)$.*
2. *Hausdorff $D$-dimensional measure of the attractor $X$ of the condensation IFS is infinite iff $D_v = D_u$.*

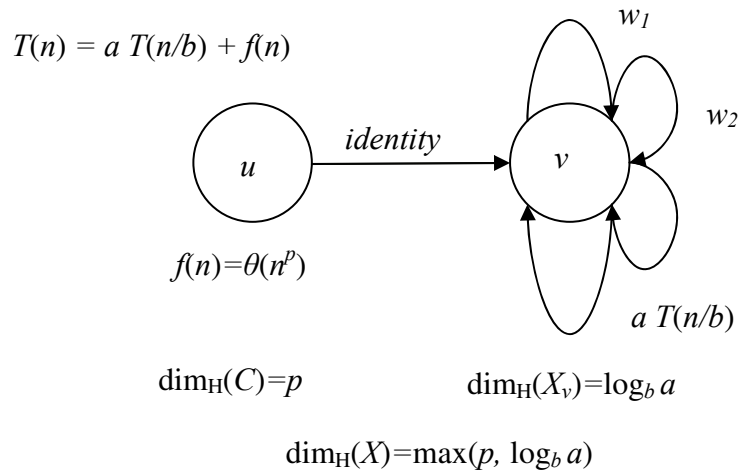

**Fig. 5.** Condensation IFS shown as a graph directed IFS. The IFS models the recurrence $T(n) = aT(n/b) + f(n)$.

For rigorous proof, see [7]. We provide some intuition behind the Theorem. Since

$$\dim_H(A \cup B) = \max\{\dim_H(A), \dim_H(B)\}$$

which holds for countable unions too, the first result follows. For the second result, it can be shown that $X$ is union of countably infinite number of fractal images, all of same dimension $D$ and same size so that the sum of their sizes does not converge and is infinite.

Mauldin-Williams theorem gives us the link between Fractal Geometry and Case 1 and Case 3 of the Master Theorem in which dimension of one image dominates the other. Just set $\epsilon = |D_v - D_u|$. For Case 2, when the dimensions are equal, it gives further insight into Hausdorff measure of the underlying fractal image.

We can now state our main result.

**Theorem 2.** *Let $T(n) = qT(n/b) + f(n)$ be a recurrence of a divide-and-conquer algorithm as in the Master Theorem. There exists a non-overlapping condensation IFS with $q$ many transformations, each of contractive factor $1/b$, whose iterations correspond to the execution of the algorithm. Let Hausdorff dimension of the attractor of* the IFS be D. If Hausdorff D-dimensional measure of the attractor is finite, then the solution of the recurrence is $T(n) = \Theta(n^D)$, else it is $T(n) = \Theta(n^D \log n)$.

The Theorem follows from Equation (3) which leads to the proof of Master Theorem in [3] and from Mauldin-Williams Theorem.

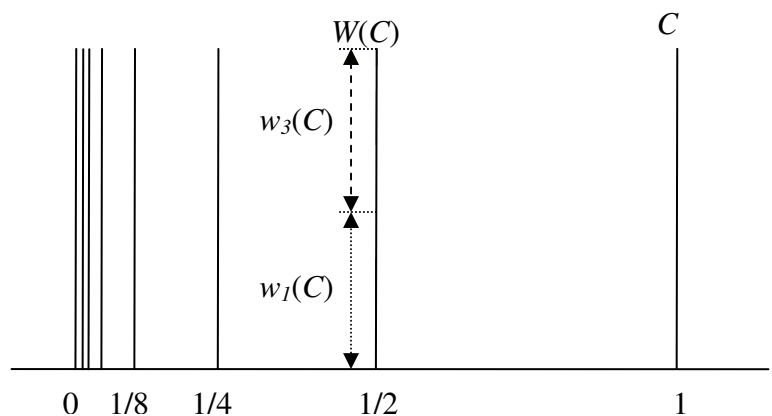

**Fig. 6.** Illustration of geometrical interpretation of Fast Fourier Transform recurrence $T(n) = 2T(n/2) + \Theta(n)$.

*Example 4.* Consider the recurrence as in Fast Fourier Transform:
$$T(n) = 2T(n/2) + \Theta(n).$$
Choose the affine transformations of the IFS to be $w_1$ and $w_3$ as shown in Figure 3 and in Example 2, and $C$ to be a vertical line segment at point (1,0) as shown in Figure 6.

Here $D_u = D_v = 1$, so 1-dimensional Hausdorff measure is infinite. This is obvious from the fact that the image in Figure 6 has infinite total length. From Theorem 2, the solution of the recurrence is $T(n) = \Theta(n \log n)$.

*Example 5.* Consider the recurrence
$$T(n) = 2T(n/2) + \Theta(n^2).$$

Refer to Figure 4. Visualize it in 3-D with the third $z$ dimension out of the page, and the tree-like image $C$ replaced by an image with dimension 2 e.g. a filled square, which is parallel to the $yz$-plane and intersects $xy$-plane at $x = 1$. We have $D_u = 2$ and $D_v = 1$. The dimension of the attractor of the condensation IFS is 2. The 2-dimensional measure is finite as the total area of the attractor is finite. From Theorem 2, $T(n) = \Theta(n^2)$.

*Example 6.* Consider the recurrence

$$T(n) = 4T(n/2) + \Theta(n^2).$$

Refer to Figure 6. Visualize it in 3-D with the third dimension out of the page, and the vertical line segment $C$ replaced by an image with dimension 2 e.g. a filled square, like in Example 5. We have $D_u = 2$ and $D_v = 2$. The dimension of the attractor of the condensation IFS is 2. The 2-dimensional measure is infinite as the total area of the attractor is infinite. From Theorem 2, solution is $T(n) = \Theta(n^2 \log n)$.

## 6   Conclusions

We provided geometrical interpretation of the Master Theorem to solve divide-and-conquer recurrences. It is natural to see the connection between the exponent $D$ in the time complexity $T(n) = \Theta(n^D)$ of such an algorithm and the Hausdorff dimension of an underlying fractal image. Another connection can be also seen. The observation

$$\Theta(n^{D_1} + n^{D_2}) = \Theta(n^{\max\{D_1, D_2\}})$$

has its fractal analog in

$$\dim_\mathrm{H}(A \cup B) = \max\{\dim_\mathrm{H}(A), \dim_\mathrm{H}(B)\}.$$

As a final remark, the results of this paper would hold for IFS in other metric spaces, e.g. in code metric space $\Sigma^\omega$ of strings over a finite alphabet $\Sigma = \{0, 1, \ldots, r\}$.